\documentclass[twocolumn,
showpacs,preprintnumbers,amsmath,amssymb]{revtex4}
\usepackage{dcolumn}
\usepackage{bm}
\usepackage[dvips]{graphicx}
\usepackage{wrapfig,subfigure}
\usepackage{amssymb}
\usepackage{color}
\def\Tr{{\rm Tr}}

\def\Re{{\rm Re}}
\def\Im{{\rm Im}}

\newcommand{\ep}{\varepsilon}

\newcommand{\dr}{^{\dagger}}

\def\<{\lesssim}
\def\>{\gtrsim}

\begin{document}

\title{Quantum Measurements of Coupled Two-Level Systems}

\author{L. Fedichkin$^{(1)}$, M. Shapiro$^{(2)}$, and M. I. Dykman$^{(1)}$\footnote{e-mail: dykman@pa.msu.edu}}
\affiliation{$^{(1)}$ Department of Physics and Astronomy,
Michigan State University, East Lansing, MI 48824\\
$^{(2)}$Department of Mathematics,
Michigan State University, East Lansing, MI 48824}

\date{\today}

\begin{abstract}
We propose an approach to measuring nonresonant coupled systems, which gives a parametrically smaller error than the conventional fast projective measurements. The approach takes into account that, due to the coupling, excitations are not entirely localized on individual systems. It combines high spectral selectivity of the detector with temporal resolution and uses the ideas of the quantum diffusion theory. The results bear on quantum computing with perpetually coupled qubits.
\end{abstract}


\maketitle

\section{Introduction}
\label{sec:Intro}

The understanding of quantum measurements has significantly advanced in recent years, in part due to the fast development of quantum information theory \cite{Clarke2008,Clerk2008}. Measurements constitute a necessary part of the operation
of a quantum computer. In the context of quantum computing, it is often implied that measurements are performed on individual two-state systems, qubits and that during measurements  qubits are isolated from each other. However, in many proposed implementations of quantum computers the qubit-qubit coupling may not be completely turned off. The interest in measuring coupled systems is by no means limited to quantum computing; however, qubits provide a convenient language for formulating the problem.

In a system of coupled qubits (coupled quantum systems) excitations are not entirely localized on  individual qubits even where the qubits have different energies. Therefore a  one-qubit measurement may miss the excitation mostly localized on the qubit which is being measured. A measurement may also give a false-positive result: the detected excitation may be mostly localized on another qubit but have a tail on the measured qubit. In the context of quantum computing, this is a significant complication, since the overall error accumulates with the number of qubits.

The most simple type of measurements which was essentially implied above is a fast von Neumann-type projective measurement \cite{Neumann}. Alternatively one can use continuous measurements, in which the signal from the qubit is accumulated over time \cite{Clerk2008}. They are often implemented as quantum non-demolition measurements (QNDMs) in which the quantity to be measured (like population of the excited state) is preserved while a conjugate quantity (like phase) is made uncertain. As we will see, the standard QNDMs do not solve the precision problem for interacting qubits.

The goal of the present paper is to find a way of measuring nonresonantly coupled qubits that gives a parametrically smaller error than standard fast or continuous one-qubit measurements. The idea is to combine temporal and spectral selectivity so as to take advantage of different time and energy scales in the system. The proposed measurement is continuous. However, it is not of a QNDM-type, both the qubit population and the phase of the qubit wave function are changed.

We will consider the situation where excitations are strongly (albeit incompletely) localized  on individual qubits, at least for the time that significantly exceeds the duration of a measurement. This can be accomplished by appropriately tuning the qubit energies $\ep_n$. For one excitation, localization of stationary states is well-understood since Anderson's work \cite{Anderson1958} on disordered systems where $\ep_n$ are random. Anderson localization requires that the bandwidth $h$ of the energies $\ep_n$ be much larger than the typical nearest-neighbor hopping integral $J$. One-excitation localization becomes stronger for the same $h/J$, i.e.,  the localization length becomes smaller if $\ep_n$ are tuned in a regular way so as to suppress resonant excitation transitions. The problem of localization of multiple excitations is far more complicated because the number of states exponentially increases with the number of excitations. However, at least for a one-dimensional qubit system, by tuning $\ep_n$ one can still obtain strong localization of all excitations for a time $\propto J^{-1}(h/J)^5$ \cite{Santos2005,Dykman2005c}; here and below we set $\hbar =1$.

We propose to detect an excitation by resonantly coupling the measured qubit (MQ) to a detecting two-level system (DS). If the MQ was initially excited and the DS was in the ground state,  the excitation can move to the DS. There its energy will be transferred to the reservoir, and the change of the state of the reservoir will be directly detected. For example, the DS can emit a photon that will be registered by a photodetector. The typical rate of photon emission $\Gamma$ should largely exceed the rate of resonant (but incoherent) excitation hopping from the MQ to the DS, so that the probability for the excitation to go back onto the MQ be small. The rate $\Gamma$ should also largely exceed the interaction-induced shift of the MQ energy levels. At the same time, $\Gamma$ should be small compared to the bandwidth of site energies $h$. Then the qubits adjacent to the MQ  are not in resonance with the DS, and the rate of excitation transfer from these qubits to the DS is small.

If the above conditions are met, there should be a time interval within which an excitation localized mostly on the MQ will be detected with large probability, whereas excitations localized mostly on neighboring qubits will have a very small probability to trigger a detection signal. As we show, the associated errors are much smaller than in a projective measurement.

State measurements for coupled qubits have been performed with Josephson-junction based systems, where there were studied oscillations of excitations between the qubits \cite{McDermott2005,Katz2006}. The oscillations occurred where the qubits were tuned in resonance with each other. The effect of the interaction on measurements in the case of detuned qubits, which is of interest for the present work, was not analyzed. The approach proposed here, which combines spectral and temporal selectivity to enable high resolution is different not only from the standard continuous quantum measurements \cite{Clerk2008} but also from other types of time-dependent quantum measurements, cf. Refs.~\onlinecite{Heiblum1996,Han2001,Korotkov2001}.

In Sec.~II we describe the system of two qubits and a resonant inelastic-scattering based DS. We identify the range of the relaxation rate of the DS and the qubit parameters where the measurement is most efficient. In Sec.~III the description of the time evolution of the system is given and it is shown that the decay rates of the stationary states of the qubits are strongly different. The detailed theory for one- and two-qubit excitations is given in Appendices~\ref{sec:AppendixA} and \ref{sec:AppendixB}, respectively. Section~IV describes how the initial states of the system can be efficiently discriminated, and the analytical results are compared with a numerical solution. Section~V contains concluding remarks, including an extension of the results to a multi-qubit system.

\section{The model}
\label{sec:model}
\subsection{A two-qubit system}

\begin{figure}[h]
\includegraphics[width=3.0in]{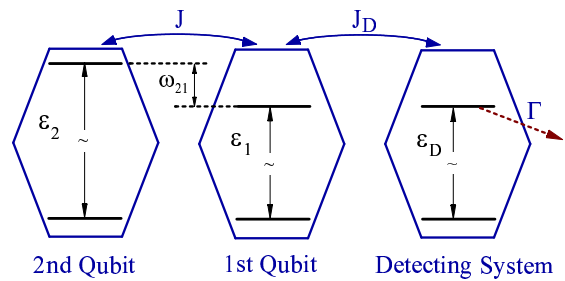}
\caption{(Color online) The measurement scheme. Qubits 1 and 2 are not in resonance, but are perpetually coupled, with the coupling constant $J\ll \omega_{21}$, where $\omega_{21}=\ep_2-\ep_1$ is the level detuning, $\omega_{21}\ll \ep_{1,2}$. The detecting system is resonantly coupled to qubit 1. When an excitation is transferred to the DS, the DS makes a transition to the ground state (for example, with photon emission) which is directly registered. The transition rate $\Gamma$ exceeds $J, J_D$, but is small compared to $\omega_{21}$ to provide spectral selectivity.}.
\label{fig:scheme}
\end{figure}

We will concentrate on a quantum measurement of two coupled two-level systems (two spin-1/2 particles or two qubits) and then discuss how the results extend to a multi-qubit system. The system is sketched in Fig.~\ref{fig:scheme}. We will assume that the qubit excitation energies (the spin Zeeman energies) $\ep_{1,2}$ largely exceed both the interaction energy $J$ and the energy difference $|\omega_{21}|$, where $\omega_{21}=\ep_2-\ep_1$; for concreteness, we assume that $\omega_{21}>0$. Via the Jordan-Wigner transformation the system can be mapped onto two spinless fermions with the Hamiltonian
\begin{eqnarray}
\label{eq:H_two_qubits}
H_S=\sum_{n=1,2}\ep_n a_n\dr a_n + \frac{1}{2}J(a_1\dr a_2+a_2\dr a_1) + J\Delta a_1\dr a_2\dr a_2 a_1.
\end{eqnarray}
Here, the subscript $n=1,2$ enumerates the coupled qubits, and $a_n, a_n\dr$ are the fermion annihilation and creation operators. The ground state corresponds to no fermions present. A fermion on site $n$ corresponds to the $n$th qubit being excited. The parameter $J$ in this language is the hopping integral, whereas $J\Delta$ describes the interaction energy of the excitations.

We will assume that $J\ll \omega_{21}$. In this case the stationary single-particle states of the system $|\psi^{(\rm st)}_{1,2}\rangle$
are strongly localized on sites 1 or 2,
\begin{eqnarray}
\label{eq:stationary_states}
|\psi^{(\rm st)}_n\rangle\approx C\left(|n\rangle - \mu(-1)^{n}|3-n\rangle\right) \quad (n=1,2),
\end{eqnarray}
where $\mu=2\delta\ep_1/J\approx -J/2\omega_{21}$ and $C=(1+\mu^2)^{-1/2}$ ($\delta\ep_1=[\omega_{21}-\sqrt{\omega_{21}^2+J^2}]/2$ is the shift of the energy level $\ep_1$ due to excitation hopping).

Measurements are done by attaching a detector to a qubit, i.e., to the physical system represented by the qubit. For concreteness, we assume that the measured qubit (MQ) is qubit 1. It is seen from Eq.~(\ref{eq:stationary_states}) that, if the measurement is fast projective and the system is in state $|\psi^{(\rm st)}_{1}\rangle$, the occupation of this state will be detected with an error $\mu^2$. With probability $\mu^2$ the detector will ``click" also if the system is in state $|\psi^{(\rm st)}_{2}\rangle$. The same error occurs, evidently, in a simple continuous measurement that does not involve energy transfer, like measurements with quantum point contacts or tunnel junctions \cite{Gurvitz1997,Korotkov2001}. Moreover, the decoherence of the qubit brought about by such a measurement, even where its rate is small compared to $\omega_{21}$, will ultimately lead to spreading of the excitation over both states $|\psi^{(\rm st)}_{1,2}\rangle$, which leads to an extra limitation on the measurement precision.

\subsection{Resonant inelastic-scattering detector}

The state of the system can be determined with a higher precision using a detector that involves inelastic transitions. A simple model is provided by a two-level DS which is resonant with qubit 1, see Fig.~\ref{fig:scheme}. The measurement is the registration of a transition of this system from its excited to the ground state; for example, it can be detection of a photon emitted in the transition.  The Hamiltonian of the qubit-DS system is
\begin{eqnarray}
\label{eq:qubit+DS_Hamiltonian}
H=H_S+\ep_Da_D\dr a_D+\frac{1}{2}J_D(a_1\dr a_D+a_D\dr a_1),
\end{eqnarray}
where $\ep_D$ is the energy of the excited state of the DS and $J_D$ characterizes the coupling of the DS to qubit 1, $J_D\ll \omega_{21}$.

The qubit-DS dynamics can be conveniently analyzed by changing to the rotating frame with a unitary transformation $U(t)=\exp[-i\ep_1t\sum_{\alpha=1,2,D}a_{\alpha}\dr a_{\alpha}]$. We assume that relaxation of the DS is due to coupling to a bosonic bath (photons). Provided this coupling is weak, the qubit-DS  dynamics in slow time (compared to $\ep_1^{-1}$) is described by a Markov equation
\begin{eqnarray}
\label{eq:QKE}
\dot\rho=i[\rho,\tilde H]-\Gamma\left(a_D\dr a_D\rho-2a_D\rho a_D\dr +\rho a_D\dr a_D\right).
\end{eqnarray}
Here, $\tilde H=H-\ep_1\sum_{\alpha=1,2,D}a_{\alpha}\dr a_{\alpha}$, and $\Gamma\equiv\Gamma(\ep_D)$ is the DS decay rate. Equation (\ref{eq:QKE}) applies provided $\Gamma\ll \ep_1$; in addition, the dispersion of $\Gamma$ has been disregarded, $|\omega_{21}(d\Gamma/d\ep_D)|\ll \Gamma$. The renormalization of $\ep_D$ due to the coupling to the bath is assumed to be incorporated into $\ep_D$, and the bath temperature $T\ll \ep_D/k_B$, so that there are no spontaneous transitions of the DS from the ground to the excited state,

Equation (\ref{eq:QKE}) should be solved with the initial condition that for $t=0$ the DS is in the ground state whereas the qubits are in a state to be measured. As a result of the excitation transfer from the qubits, the DS can be excited and then it will make a transition to the ground state. The directly measured quantity is the probability $R(t)$ that such a transition has occurred by time $t$,
\begin{equation}
\label{eq:R_probability}
R(t)=2\Gamma\int\nolimits_0^t dt\,\Tr \left[\rho(t)a_D\dr a_D\right].
\end{equation}
It is clear, in particular, that if one of the qubits is excited, the excitation will be ultimately fully transferred to the DS and then further transferred to the photon bath, so that $R(t)\to 1$ for $t\to\infty$. If on the other hand both qubits are in the ground state, then $R(t)=0$.

Qubit measurements can be efficiently done for
\begin{equation}
\label{eq:DS_conditions}
\omega_{21}\gg \Gamma\gg J,J_D, |\ep_D-\ep_1|.
\end{equation}
In this case, there are no oscillations of excitations between qubit 1 and the DS. The energy detuning $\ep_D-\ep_1$ plays no role, and without loss of generality we can set $\ep_D=\ep_1$. We will start the analysis with the case where there is no more than one excitation on the qubits for $t=0$. Then one can replace $\tilde H$ with a single-excitation Hamiltonian,
\begin{equation}
\label{eq:RWA_Hamiltonian}
\tilde H\Rightarrow\omega_{21}a_2\dr a_2 +\frac{1}{2}\left(Ja_1\dr a_2 + J_Da_1\dr a_D + {\rm H.c.}\right)
\end{equation}
As we show, for a DS that satisfies conditions (\ref{eq:DS_conditions}) time evolution of $R(t)$ is characterized by two strongly different scales, which makes it possible to differentiate between the qubit states.

\section{Time evolution of the density matrix}
\label{sec:time_evolution}
\subsection{One excitation}

The continuous measurement (\ref{eq:R_probability}) requires finding expectation values $\langle a_{\beta}\dr(t) a_{\alpha}(t)\rangle \equiv \Tr a_{\beta}\dr a_{\alpha}\rho(t) \equiv \rho_{\alpha\beta}(t)$, where $\alpha,\beta$ run through the subscripts $1,2,D$. The matrix elements $\rho_{\alpha\beta} = \rho^*_{\beta\alpha}$ satisfy a system of nine linear equations that follow from the operator equations (\ref{eq:QKE}), (\ref{eq:RWA_Hamiltonian}). This system of equations is closed, the matrix elements $\rho_{\alpha\beta}$ do not mix with the expectation values $\langle a_{\alpha}(t)\rangle, \langle a_{\alpha}\dr(t)\rangle$. In the case where there is only one excitation (one fermion) they also do not mix with expectation values $\langle a_{\alpha}\dr(t)a_{\beta}\dr(t)a_{\gamma}(t)a_{\delta}(t)\rangle$, which are all equal to zero.

The solution of the equations for $\rho_{\alpha\beta}(t)$ and the analysis of the signal $R(t)$  are simplified in the range (\ref{eq:DS_conditions}). The relaxation rate of the matrix elements $\rho_{D\alpha  }$ that involve the DS is $ \Gamma$. For $\Gamma \gg J_D,J$ this rate is faster than other relaxation rates, as explained below (see Appendix A for details), and therefore over time $\Gamma^{-1}$  the matrix elements  $\rho_{D\alpha  }$ reach their quasi-stationary values. Relaxation of $\rho_{11}$, on the other hand, is determined by the excitation transfer from site 1 to the DS. This transfer is similar to quantum diffusion of defects in solids within narrow bands for weak coupling to phonons \cite{Kagan1974} or between closely spaced discrete non-stationary states of weakly coupled reorienting defects \cite{Dykman1978b}. It is characterized by rate
\begin{equation}
\label{eq:relax_rate_1}
W_1=J_D^2/2\Gamma,\qquad W_1\ll \Gamma.
\end{equation}
Equation (\ref{eq:relax_rate_1}) can be readily understood in terms of the Fermi golden rule: this is a transition rate from qubit 1 to the DS induced by the interaction $\propto J_D$, with $\Gamma$ being the characteristic bandwidth of the final states and $\Gamma^{-1}$ being the density of states at the band center, respectively. The rate $W_1$ describes relaxation of an excitation localized initially in state $|\psi^{\rm (st)}_1\rangle$.

If the excitation is localized mostly on qubit 2 and has energy $\approx \omega_{21}$, its decay rate becomes much smaller. The decay can be thought of as due to the interaction of qubit 2 with the DS, which is mediated (nonresonantly)  by qubit 1. Therefore the effective interaction energy is $\approx J J_D/2\omega_{21}$. The effective density of final states is determined by the tail of the density of states of the DS at frequency $\omega_{21}\gg \Gamma$. For exponential in time relaxation described by  Eq.~(\ref{eq:QKE}) the density of states is Lorentzian, and on the tail at frequency $\omega_{21}$ it is $\sim \Gamma/\omega_{21}^2$. Therefore the expected decay rate $W_2$ is
\begin{equation}
\label{eq:relax_rate_2}
W_2=J^2J_D^2\Gamma/8\omega_{21}^4, \qquad W_2\ll W_1.
\end{equation}

It follows from the result of Appendix~\ref{sec:AppendixA} that, if for $t=0$ the two-qubit system is in a stationary state $|\psi^{\rm (st)}_n\rangle$ given by Eq.~(\ref{eq:stationary_states}) ($n=1,2$), the probability $R_n(t)$ to receive a signal by time $t$ is
\begin{equation}
\label{eq:expon_signal_st_states}
R_n(t)=1-\exp(-W_nt) \qquad (n=1,2).
\end{equation}
The strong difference between $W_{1,2}$ and $\Gamma$ justifies the assumption that $\rho_{D\alpha}$ reaches a quasistationary value before the populations $\rho_{11}, \rho_{22}$ change. The full adiabatic theory of qubit relaxation is described in Appendix A. We note that the analysis takes into account corrections $\sim (J/\omega_{21})^2$, which are important, since they determine the difference between a fast projective measurement and the continuous measurement discussed here.

\subsection{Two excitations}

The above analysis can be readily extended to the case where both qubits 1 and 2 are initially in the excited state, i.e., initially there are two spinless fermions on sites 1 and 2. We will assume that the interaction between the excitations (fermions) $J\Delta$ is not strong, so that $|J\Delta|\ll \Gamma$. The qualitative picture of the dynamics of the system then is simple. First, over time $\sim\Gamma^{-1}$ there is established a quasi-stationary ``quantum diffusion current" from qubit 1 to the DS, which is determined by the transition rate $W_1$ and is equal to $W_1\rho_{11}(t)$. It drains stationary state $|\psi^{\rm (st)}_1\rangle$ over time $\sim W_1^{-1}$. The presence of an excitation (fermion) in state $|\psi^{\rm (st)}_2\rangle$ only weakly affects this current because of the large energy difference of the states $\omega_{21}$. After state $|\psi^{\rm (st)}_1\rangle$ is emptied, further evolution corresponds to the single-excitation decay of state $|\psi^{\rm (st)}_2\rangle$. Therefore the overall signal should be
\begin{equation}
\label{eq:sign_two_excitns}
R^{(2e)}(t)=R_1(t)+R_2(t)=2-e^{-W_1t} - e^{-W_2t}.
\end{equation}
A detailed derivation of this expression is given in Appendix~\ref{sec:AppendixB}. It is clear from the above analysis that the most interesting problem is to distinguish which of the stationary states $|\psi^{\rm (st)}_{1,2}\rangle$ is initially occupied; the situation where both of them are occupied is simpler.

\section{Resolving one-excitation states from time-dependent measurements}
\label{sec:state_resolution}

\begin{figure}[h]
\hspace*{-0.15in}\includegraphics[width=2.7in]{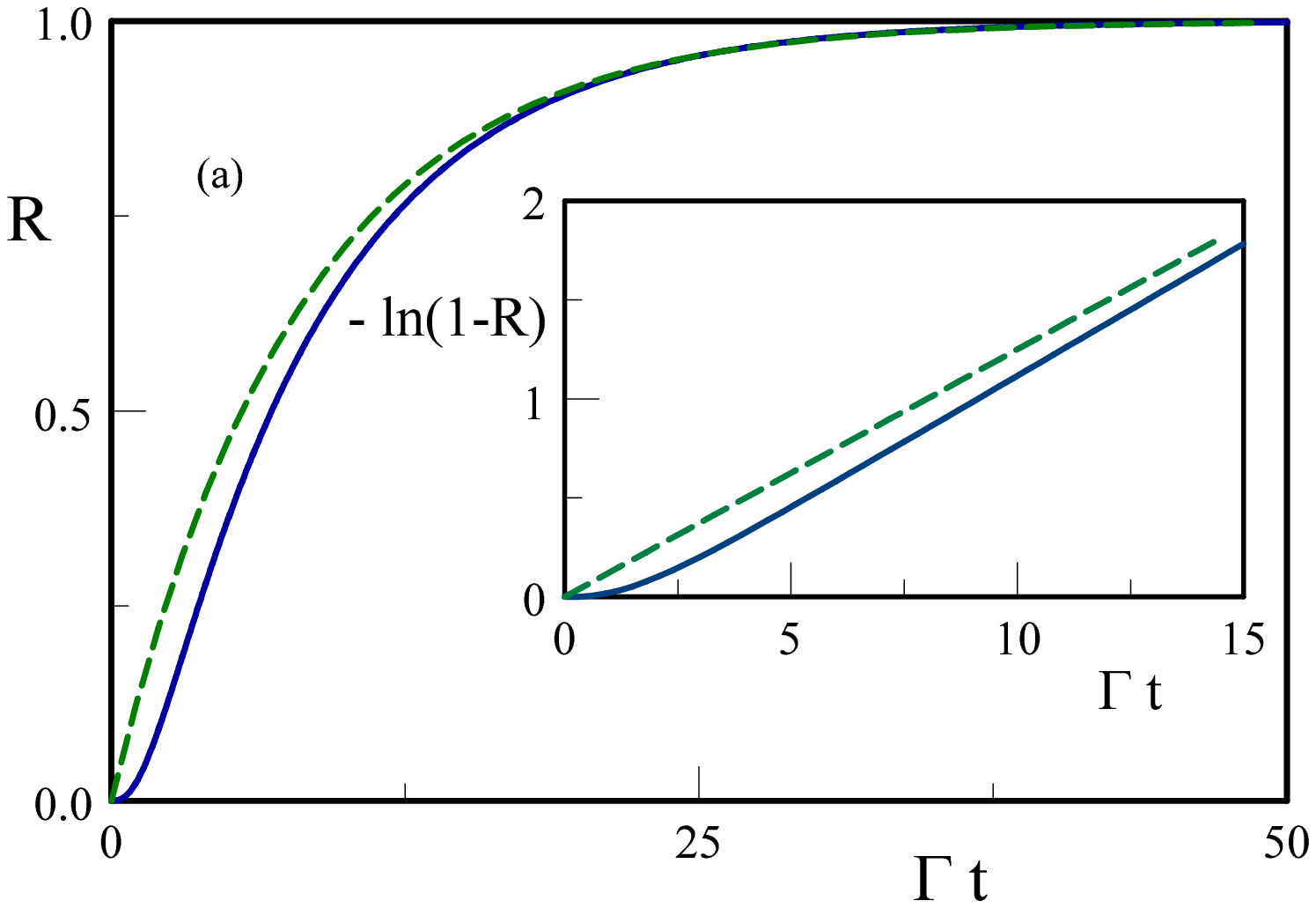}
\includegraphics[width=2.8in]{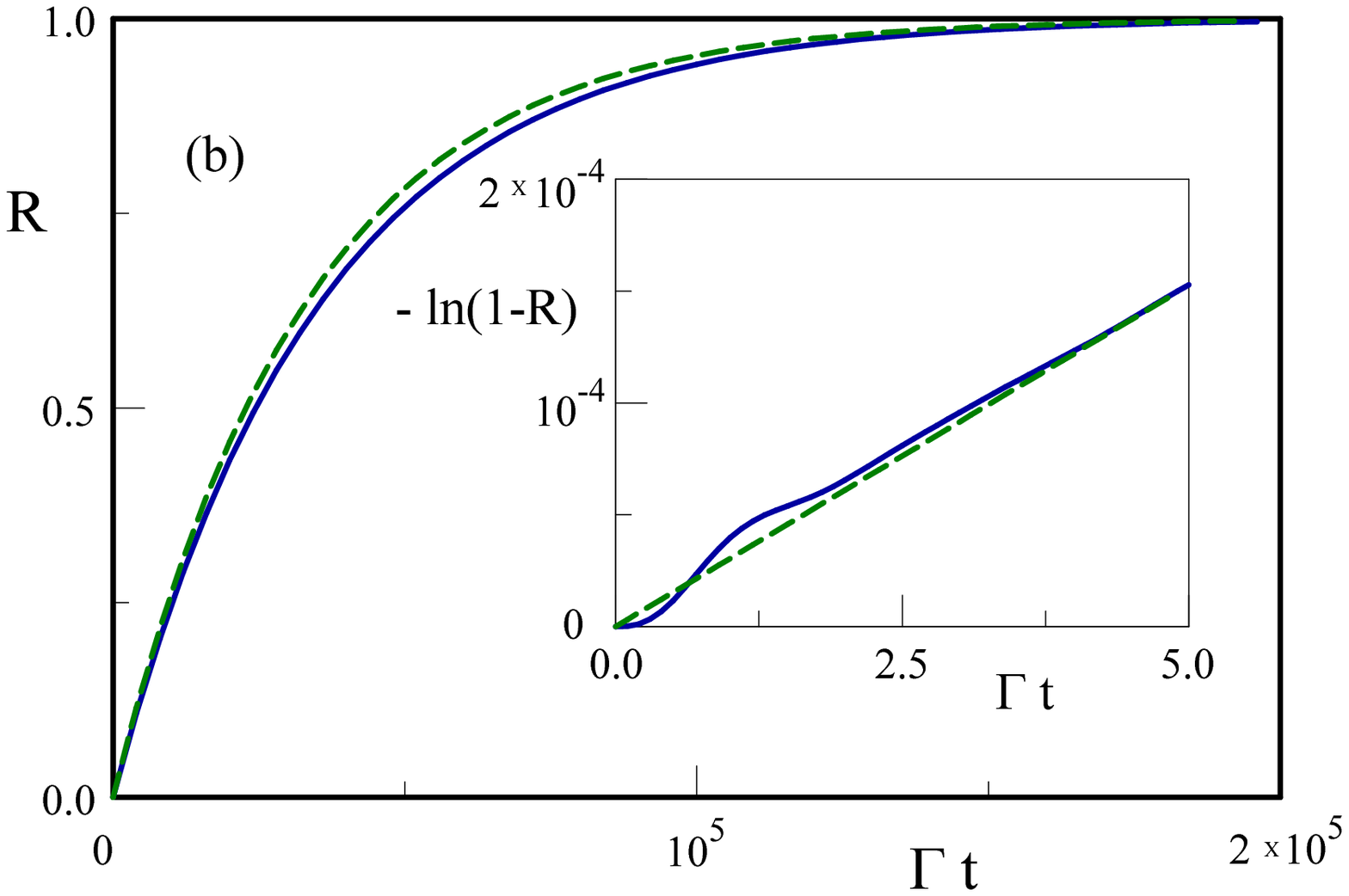}
\caption{(Color online) Time dependence of the probability to detect a signal for the two-qubit system being initially in the stationary states $|\psi^{\rm (st)}_1\rangle$ (a) and $|\psi^{\rm (st)}_2\rangle$ (b). Note the difference of time scales. The parameters are $\omega_{21}/\Gamma=4, J/\Gamma=J_D/\Gamma=1/2$.  The solid and dashed lines show, respectively, the numerical solution of the master equation and the asymptotic expressions (\ref{eq:expon_signal_st_states}) for the limiting case $\omega_{21}\gg \Gamma \gg J, J_D$. The insets show $|\log[1-R_1(t)]|$  for small time, where the difference between the numerical and asymptotic expressions is most pronounced.}
\label{fig:eigenstates_decay}
\end{figure}

\begin{figure}[h]
\includegraphics[width=2.8in]{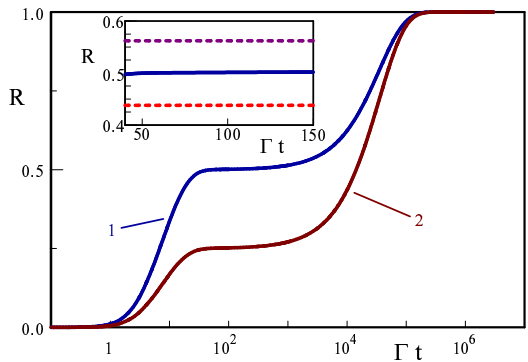}
\caption{(Color online) Time dependence of the probability to detect a signal for the two-qubit system being initially in a superposition of stationary states $|\psi(0)\rangle =\cos\theta|\psi^{\rm (st)}_1\rangle +\sin\theta|\psi^{\rm (st)}_2\rangle$. Curves 1 and 2 present the numerical solution of the master equation and refer to $\theta= \pi/4$ and $\pi/3$, respectively. The parameters of the qubits and the DS are the same as in Fig.~\ref{fig:eigenstates_decay}. Inset: the solid line shows $R(t)$ for $\theta= \pm \pi/4$ ($P_1=1/2$) in the optimal time range (\ref{eq:time_discrimination_inerval}), the dashed lines show the results of fast projective measurements on qubit 1, $|\langle 1|\psi(0)\rangle|^2$, for $\theta=\pm \pi/4$.}.
\label{fig:mixedstate_decay}
\end{figure}

The rates $W_1$ and $W_2$ of signal accumulation for different initially occupied one-excitation states are parametrically different. This enables efficient distinction between the states using a resonant DS. The results for time evolution of the signals $R (t)$ obtained by a numerical solution of the system of equations for the matrix elements $\rho_{\alpha\beta}$ are shown in Figs.~\ref{fig:eigenstates_decay} and \ref{fig:mixedstate_decay}. The figures refer, respectively, to the cases where the system of qubits is initially in one of the eigenstates $|\psi^{\rm (st)}_{1,2}\rangle$ and in a mixed state. It is seen from the figures that, even where the parameter ratios $\Gamma/\omega_{21}$ and $J/\Gamma, J_D/\Gamma$ are not particularly small, the results are well described by the asymptotic expressions (\ref{eq:expon_signal_st_states}).

Expressions (\ref{eq:expon_signal_st_states}) are obtained in the adiabatic approximation and do not describe how the matrix elements $\rho_{\alpha\beta}$ reach their adiabatic values. In particular, they do not describe the evolution of $R$ for $t\lesssim \Gamma^{-1}$. Since the DS is initially in the ground state, it follows from Eqs.~(\ref{eq:QKE}) and (\ref{eq:R_probability}) that $R(t)\propto t^2$ for $t\to 0$ in contrast to $R\propto t$ as predicted by Eq.~(\ref{eq:expon_signal_st_states}).

Breaking of the adiabaticity for short times explains the shifts of the asymptotic curves (\ref{eq:expon_signal_st_states}) with respect to the numerically calculated curves in Fig.~\ref{fig:eigenstates_decay}. A simple estimate for the shift can be obtained for detection of state $|\psi^{\rm (st)}_1\rangle$ . Here, adiabaticity is established over time $t\sim \Gamma^{-1}$; this is the relaxation time of the matrix elements $\rho_{\alpha D}, \rho_{D\alpha}$ that directly involve the DS. Then from Eq.~(\ref{eq:expon_signal_st_states}) the shift should be $\sim W_1/\Gamma$, which agrees with Fig.~\ref{fig:eigenstates_decay}(a). For the initially occupied state $|\psi^{\rm (st)}_2\rangle$ the exponential decay of $R_2$, Eq.~(\ref{eq:expon_signal_st_states}), is obtained in the double-adiabatic approximation, which exploits the interrelation between the relaxation rates $\Gamma \gg W_1\gg W_2$. The adiabatic regime is formed over time $\sim W_1^{-1}$, and the shift of the numerical curve with respect to the asymptotic one is $\sim W_2/W_1$, which agrees with Fig.~\ref{fig:eigenstates_decay} (b).

The vast difference of the time scales over which the signal accumulates depending on the initially occupied state allows one to discriminate between the states with high precision. If the initial state of the system $|\psi(0)\rangle$ is a superposition of stationary states $|\psi^{\rm (st)}_{1}\rangle$ and $|\psi^{\rm (st)}_{2}\rangle$, for $\Gamma t\gg 1$ the signal $R(t)$ is the appropriately weighted superposition of the signals $R_1(t)$ and $R_2(t)$,
\begin{eqnarray}
\label{eq:state_mixture}
&&R(t)\approx P_1\left[1-\exp(-W_1t)\right] + P_2\left[1-\exp(-W_2t)\right], \nonumber\\
&&P_n=\left|\langle \psi(0)|\psi^{\rm (st)}_n\rangle\right|^2.
\end{eqnarray}
Here, $P_{1,2}$ are the initial populations of the stationary states. We note that $P_1+P_2\neq 1$, in the general case. The initial state of the two-qubit system can be a superposition of the ground and one- or two-excitation states.

Over time $\sim W_1^{-1}$ the function $R(t)$ approaches the population $P_1$ of state $|\psi^{\rm (st)}_1\rangle$. The further change of $R(t)$ occurs over a much longer time $\sim W_2^{-1}$. This is seen in Fig.~\ref{fig:mixedstate_decay}. As explained in Appendix~\ref{sec:AppendixA}, a contribution to $R(t)$ from the fast-oscillating [as $\exp(\pm i\omega_{21}t)]$ terms in the density matrix is small compared to $(J/\omega_{21})^2$: for $\Gamma t\gg 1$ the corresponding correction to Eq.~(\ref{eq:state_mixture}) is $ \lesssim JJ_D^2/4\omega_{21}^3\Re \rho_{12}(0)\ll J^2/4\omega_{21}^2$  (we assume $J_D\sim J$).

From Eq.~(\ref{eq:state_mixture}), in a broad time interval the error in the measured population of state $|\psi^{\rm (st)}_1\rangle$ is smaller than in a fast projective measurement. If the initial state is $\psi(0)\rangle = P_1^{1/2}|\psi^{\rm (st)}_1\rangle + P_2^{1/2}\exp(i\phi)|\psi^{\rm (st)}_2\rangle$, a fast projective measurement on qubit 1 gives
\begin{eqnarray}
\label{eq:fast_measurement}
|\langle 1|\psi(0)\rangle|^2\approx \left(1-\frac{J^2}{4\omega_{21}^2}\right)\left|P_1^{1/2} + P_2^{1/2}e^{i\phi}\frac{J}{2\omega_{21}}\right|^2.
\end{eqnarray}
This differs from $P_1$ by $\sim J/\omega_{21}$ for $P_1\sim P_2$; in the case of strongly different populations where  $P_1/P_2$ or $P_2/P_1$ is $\lesssim J^2/\omega_{21}^2$ the difference becomes $\sim J^2/\omega_{21}^2$. The proposed measurement gives a much smaller (in fact, a parametrically smaller) error. We have
\begin{equation}
\label{eq:time_discrimination_inerval}
|R(t)-P_1|\ll \frac{J^2}{4\omega_{21}^2} \quad {\rm for} \quad e^{-W_1t},W_2t\ll \frac{J^2}{4\omega_{21}^2}.
\end{equation}
In the explicit form, the time interval for high-accuracy measurement of state $|\psi^{\rm (st)}_1\rangle$ is determined by condition $2\ln(2\omega_{21}/J)\ll W_1t\ll (\omega_{21}/\Gamma)^2$. This condition is easy to satisfy in the range (\ref{eq:DS_conditions}). We emphasize that, as seen from Eq.~(\ref{eq:fast_measurement}), in most cases the error of a fast projective measurement is parametrically larger than $J^2/\omega_{21}^2$.  The difference between the proposed approach and a fast projective measurement is illustrated in the inset in Fig.~\ref{fig:mixedstate_decay}.

Advantageous features of the proposed approach can be seen from a comparison with a similar but a seemingly simpler scheme in which one directly turns on the coupling of qubit 1 to a thermal reservoir and detects the emitted excitation. This scheme is analogous to the scheme implemented in Josephson phase qubits \cite{McDermott2005,Katz2006} where decay of one of the qubits was effectively ``turned on" by reducing the appropriate tunnel barrier. In such a scheme the decay rate of the initial stationary state localized mostly on qubit 1 is determined by the introduced relaxation rate of this qubit $\Gamma$. The decay rate of the stationary state localized mostly on qubit 2 can be estimated, following the arguments of Sec.~\ref{sec:time_evolution}, as $\sim J^2\Gamma/\omega_{21}^2$, provided $\omega_{21}\gg \Gamma, J$. Therefore, if the system is initially in state $|\psi^{\rm (st)}_2\rangle$, over the lifetime $\Gamma^{-1}$ of state $|\psi^{\rm (st)}_1\rangle$ the detector will ``click" with probability $\sim (J/\omega_{21})^2\Gamma t\gtrsim (J/\omega_{21})^2$. As a result, there is no gain compared to a fast projective measurement.

\section{Conclusions}
\label{sec:conclusions}

In this paper we have demonstrated that a state of two coupled two-level systems (qubits), which are detuned from each other, can be determined with an accuracy much higher than that of a fast projective measurement. This is accomplished by combining frequency and temporal selectivity. The detector is tuned in resonance with the qubit to which it is directly coupled. When the detector relaxation rate $\Gamma$ is higher than the coupling $J, J_D$ but less than the energy difference between the qubits $\omega_{21}$, the overall relaxation of the system is characterized by two strongly different time scales. One relaxation rate, $W_1= J_D^2/2\Gamma$, determines the decay of the excited stationary state localized mostly on the resonant qubit. The decay rate of the excited stationary state localized mostly on the non-resonant qubit $W_2=(JJ_D)^2\Gamma/8\omega_{21}^4$ is parametrically smaller. This makes it possible to discriminate between the states using time-dependent measurements.

The measurement requires turning on the interaction between the measured qubit and the DS. Alternatively, and often more conveniently, one can tune the DS in resonance with the qubit. The required precision is determined by the decay rate of the DS $\Gamma$. Moreover, one can slowly sweep the DS energy level through the qubit energy, so that the energies stay in resonance (to accuracy $\sim\Gamma$) for a time $\gtrsim W_1^{-1}$.

The results can be generalized to the case of a many-qubit system. Of particular interest is a qubit chain with Hamiltonian
\begin{eqnarray}
\label{eq:chain_hamiltonian}
H=&&\sum_n\ep_na_n\dr a_n + \frac{1}{2}J\sum_n(a_n\dr a_{n+1}+a_{n+1}\dr a_n)\nonumber\\
&&+ J\Delta \sum_na_n\dr a_{n+1}\dr a_{n+1} a_n,
\end{eqnarray}
which is an immediate extension of the two-qubit Hamiltonian (\ref{eq:H_two_qubits}). As mentioned in Sec.~I, for appropriately tuned site energies $\ep_n$ all many-particle excitations in such a chain remain localized for a long time, which scales as a high power of $h/J$ ($h$ is the typical bandwidth of energies $\ep_n$) \cite{Santos2005,Dykman2005c}. However, excitations are not fully localized on individual sites, in (quasi)stationary states the tails of their wave functions on neighboring sites are $~\sim J/h$.

Following the proposed method, to determine whether there is an excitation localized in a quasistationary state centered at a given qubit one should couple this qubit resonantly to a DS. The excitation will be detected over time $\sim W_1^{-1}$. Excitations localized mostly on neighboring qubits will not affect the measurement as long as the duration of the measurement is small compared to $W_2^{-1}$. The interaction between excitations will not affect this result for appropriately chosen decay rate of the DS. The reduction of the measurement error is an important prerequisite for scalable quantum computing with perpetually coupled qubits.

\appendix

\section{The adiabatic approximation}
\label{sec:AppendixA}

In the presence of one excitation on the qubits, the dynamics of the two-qubit system coupled to the DS is described by Eqs.~(\ref{eq:QKE}), (\ref{eq:RWA_Hamiltonian}), which can be written as nine linear first-order equations  for the matrix elements $\rho_{\alpha\beta}(t)=\langle a_{\beta}(t)\dr a_{\alpha}(t)\rangle$. Formally one can solve these equations  by finding the corresponding eigenvalues and eigenfunctions. For a strong inequality between the transition frequency $\omega_{21}$, on the one hand and, on the other hand, the decay rate $\Gamma$ and the hopping integrals $J, J_D$ (\ref{eq:DS_conditions}), the eigenvalues can be separated into those corresponding to fast weakly damped oscillations and to slow evolution. There are 4 ``fast" eigenvalues with imaginary part close to $\pm \omega_{21}$ . In particular, as seen from Eq.~(\ref{eq:QKE}), to zeroth order in $J,J_D$ we have $\rho_{12},\rho_{D2}\propto\exp(i\omega_{21}t)$. In turn, where $\Gamma \gg J, J_D$ the ``slow" eigenvalues can be separated into three eigenvalues with real parts $\propto -\Gamma$. In particular, to zeroth order in $J, J_D$ we have $\rho_{DD}\propto \exp(-2\Gamma t)$ and $\rho_{D1}=\rho_{1D}^* \propto \exp(-\Gamma t)$ . Then there are two more ``slow" eigenvalues, which as we will show are given by Eqs.~(\ref{eq:relax_rate_1}) and (\ref{eq:relax_rate_2}).

We start with the effect of slow motion on the measured signal $R(t)$. The analysis of slow dynamics can be done in the adiabatic approximation. The relaxation rate of the matrix elements $\rho_{D\alpha}$ ($\alpha=1,2,D$) is $\propto \Gamma$. Therefore over time $t\gg \Gamma^{-1}$ these matrix elements reach quasi-stationary values. These values adiabatically follow the slowly evolving matrix elements $\rho_{nm}$, where Roman subscripts $n,m$ enumerate the excited states of the qubits. In particular, by noting that $(\partial/\partial t)\sum_{\alpha}\rho_{\alpha\alpha}=-2\Gamma\rho_{DD}$ we obtain
\begin{equation}
\label{eq:rhoDD_vs_rho11_22}
\rho_{DD} \approx - (\dot\rho_{11} + \dot\rho_{22})/2\Gamma.
 \end{equation}

\subsection{Evolution of the one-excitation stationary state resonant with the DS}

 We assume first that the system is in the stationary state $|\psi^{\rm (st)}_1\rangle$, which has energy close to the DS energy. In this case, for $t=0$ we have $\rho_{11}\gg \rho_{22}\approx (J/\omega_{21})^2\rho_{11}$ and all off-diagonal matrix elements $\rho_{\alpha\beta}$ with $\alpha\neq \beta$ are small compared to $\rho_{11}$. As we will see, this hierarchy persists for $t>0$ as well. For $t\gg \Gamma^{-1}$  to leading order in $J, J_D$ we have $\rho_{D1}\approx i(J_D/2\Gamma)(\rho_{DD}-\rho_{11})+i(J/2\Gamma)\rho_{D2}$ and $\rho_{DD}\approx -(J_D/2\Gamma)\Im\rho_{D1}$. Substituting $\rho_{D1}$ into the equations for $\dot\rho_{11}$ and eliminating $\rho_{DD}$ we obtain, to leading order,
 \begin{eqnarray}
 \label{eq:rho11rate}
 &&\dot\rho_{11}\approx -W_1\rho_{11}+\Delta_{11}, \\ &&\Delta_{11}=\frac{1}{2}iJ(\rho_{12}-\rho_{21})+\frac{JJ_D}{2\Gamma}\Re\rho_{D2}, \nonumber
\end{eqnarray}
where the relaxation rate $W_1$ is given by Eq.~(\ref{eq:relax_rate_1}). In agreement with the adiabaticity assumption, we have $W_1\ll \Gamma$. The term $\Delta_{11}$ does not affect the evolution of $\rho_{11}$ if the system is initially in state $|\psi^{\rm (st)}_1\rangle$, but it is important for the analysis of evolution from state $|\psi^{\rm (st)}_2\rangle$.

To find time evolution of $\rho_{22}$ one has to use equations
\begin{eqnarray}
\label{eq:rho12_and_rho22}
&&\dot\rho_{12}=i\omega_{21}\rho_{12}+\frac{1}{2}iJ(\rho_{11}-\rho_{22})-\frac{1}{2}iJ_D\rho_{D2},\quad\\
&&\dot\rho_{22} = - \frac{1}{2}iJ(\rho_{12}-\rho_{21})\nonumber
\end{eqnarray}
that immediately follow from operator equation (\ref{eq:QKE}). Using the similar equation for $\dot\rho_{D2}$ and the expression for $\rho_{D1}$ given above one can show that, for $\rho_{11}\gg \rho_{22}$, $\rho_{D2}\approx i(JJ_D/4\omega_{21}\Gamma)\rho_{11}$. This expression should be substituted into Eq.~(\ref{eq:rho12_and_rho22}) for $\rho_{12}$. One than obtain $\Im\rho_{12}\approx (J/4\omega_{21}^2)\dot\rho_{11}$. Then from the second equation (\ref{eq:rho12_and_rho22} $\dot\rho_{22}\approx (J/2\omega_{21})^2\dot\rho_{11}$. Therefore, if the system is initially in the stationary state  $|\psi^{\rm (st)}_1\rangle$, the relation between $\rho_{22}(t)$ and $\rho_{11}(t)$ remains unchanged, to leading order in $J$.

From Eqs.~(\ref{eq:R_probability}), (\ref{eq:rhoDD_vs_rho11_22}), (\ref{eq:rho11rate}) we see that, if the two-qubit system is initially in the stationary state $|\psi^{\rm (st)}_1\rangle$, the probability to have detected this state by time $t$ is
$R_1(t)=\exp\left(-W_1t\right)$. We use this expression in Eq.~(\ref{eq:expon_signal_st_states}).

\subsection{Evolution of the nonresonant one-excitation stationary state}

If the system is initially in state $|\psi^{\rm (st)}_2\rangle$, we have $\rho_{11}\ll \rho_{22}$. Then it is important to keep track of higher-order corrections in $J/\omega_{21}$ in the equations for $\rho_{\alpha\beta}$. In particular, in the adiabatic approximation one has to write $\rho_{D2}(t)$ as
\begin{equation}
\label{eq:rho_D2}
\rho_{D2}(t) \approx [2(\omega_{21} + i\Gamma)]^{-1}\left[J_D\rho_{12}(t) + i\frac{JJ_D}{2\Gamma}\rho_{11}(t)\right]
\end{equation}

The decay rate of $\rho_{12}(t)$ can be obtained by substituting Eq.~(\ref{eq:rho_D2}) into Eq.~(\ref{eq:rho12_and_rho22}) for $\dot\rho_{12}$, which gives for the decay rate an expression $(J_D/2\omega_{21})^2\Gamma$. This rate is small compared to the decay rate $\Gamma$ of $\rho_{D2}$, justifying the adiabatic approximation used in Eq.~(\ref{eq:rho_D2}). At the same time, it is large compared to the decay rate of $\rho_{22}$. To describe decay of $\rho_{22}$ , we substitute the adiabatic solution of Eqs.~(\ref{eq:rho12_and_rho22}) for $\Im\rho_{12}$ [with account taken of (\ref{eq:rho_D2})] into Eq.~(\ref{eq:rho12_and_rho22}) for $\rho_{22}$. This gives
\begin{equation}
\label{eq:decay_rho22}
\dot\rho_{22}\approx -W_2\rho_{22},
\end{equation}
with the decay rate $W_2$ of the form of Eq.~(\ref{eq:relax_rate_2}).

Noting that $\Re\rho_{12}\approx (J/2\omega_{21})\rho_{22}$, we obtain from Eqs.~(\ref{eq:rho11rate}), (\ref{eq:rho_D2}) that, for time $t\gg W_1^{-1}$, we have $\rho_{11}(t)\approx (J/2\omega_{21})^2\rho_{22}(t)$. Then from Eqs.~(\ref{eq:R_probability}, (\ref{eq:rhoDD_vs_rho11_22}), and (\ref{eq:decay_rho22}) we obtain the explicit expression (\ref{eq:expon_signal_st_states}) for the probability to detect state $|\psi^{\rm (st)}_2\rangle$ in time $t$.

If initially the system is in a superposition of the states $|\psi^{\rm (st)}_{1,2}\rangle$, along with slowly varying (on time $\omega_{21}^{-1}$) solutions there emerge terms in the density matrix that contain fast oscillating factors $\propto \exp(\pm i\omega_{21}t)$.  One can show that, for $t\gtrsim\Gamma^{-1}$, their decay is controlled by the decay of the fast-oscillating component $\rho_{12}^{\rm (f)}(t)=[\rho_{21}^{\rm (f)}(t)]^*$, which is characterized by the factor $\exp(-W_1t/2)$.  The corresponding fast-oscillating term in the population of the excited state of the DS is $\rho_{DD}^{\rm (f))}(t)\sim i(JJ_D^2/8\Gamma\omega_{21}^2)\rho_{12}^{\rm (f)}(t)+{\rm c.c.}$. Therefore the contribution of the fast-oscillating terms to the observed probability $R(t)$ is $\lesssim (JJ_D^2/2\omega_{21}^3 )\Re \rho_{12}(0)$. It is small compared to the measurement error $\sim J^2/\omega_{21}^2$ that we want to overcome.

\section{Dynamics of the two-excitation state}
\label{sec:AppendixB}

In the presence of two excitations, along with the two-particle matrix elements $\rho_{\alpha\beta}(t) =\langle a_{\beta}\dr(t)a_{\alpha}(t)\rangle$ the dynamics of the system is characterized by the four-particle matrix elements $\rho_{\alpha\beta\gamma\delta}(t)=\langle a_{\delta}\dr(t)a_{\gamma}\dr(t)a_{\beta}(t)a_{\alpha}(t)\rangle$. For a two-qubit system coupled to a DS each Greek subscript runs through 3 values, and the dynamics is described by a set of 18 linear differential equations that follow from the equation of motion for the density matrix in the operator form (\ref{eq:QKE}): 9 equations for $\dot\rho_{\alpha\beta}$ and 9 equations for $\dot\rho_{\alpha\beta\gamma\delta}$. The number of independent matrix elements follows from the permutation symmetry $\delta\rightleftharpoons \gamma,\, \beta\rightleftharpoons \alpha$ (a permutation is accompanied by a sign change) and the commutation relations for the fermion operators $a_{\alpha}, a_{\alpha}\dr$. The system of equations for the four-particle matrix elements is closed: $\rho_{\alpha\beta\gamma\delta}$ are expressed in terms of each other, whereas their evolution affects the two-particle matrix elements $\rho_{\alpha\beta}$.

We will again use the adiabatic approximation to study the dynamics and will start with the matrix elements $\rho_{\alpha\beta\gamma\delta}$. Formally, their time evolution is determined by the eigenvalues of the system of equations for $\rho$. Four of these eigenvalues correspond to fast oscillations with imaginary part $\approx \pm\omega_{21}$. In the limit $J\to 0$ the corresponding eigenvectors $\rho_{\alpha\beta\gamma\delta}$ have one of the subscripts equal to 2, whereas other three subscripts are two 1 and one $D$ or two $D$ and one 1. Among the eigenvalues that describe ``slow" motion there are two $\approx -\Gamma$ and two $\approx -2\Gamma$. In the limit $J\to 0$ the corresponding eigenvectors are $\rho_{D212},\rho_{12D2}$ and $\rho_{nnDD}$ ($n=1,2$). Only one eigenvector controls the evolution of the four-particle matrix elements on times longer than $\Gamma^{-1}$, the leading term in this eigenvector is $\rho_{1221}$.

Over time $t\gtrsim \Gamma^{-1}$ the matrix elements $\rho_{\alpha\beta\gamma\delta}$ reach their quasi-stationary values which are determined by $\rho_{1221}(t)$. They can be found by disregarding $\dot\rho_{\alpha\beta\gamma\delta}$ in all equations except for the equation for $\dot\rho_{1221}$. This gives, in particular,
\begin{eqnarray}
\label{eq:4_corrltrs}
\rho_{D221}(t)\approx -\frac{iJ_D}{2\Gamma}\rho_{1221}(t),\quad \dot\rho_{1221}\approx - W_1\rho_{1221},
\end{eqnarray}
whereas $\rho_{1D21}\approx i(JJ_D/4\omega_{21}\Gamma)\rho_{1221}$. It is seen from Eq.~(\ref{eq:4_corrltrs}) that all slowly varying components of $\rho_{\alpha\beta\gamma\delta}$ decay to zero over time $\sim W_1^{-1}$. This makes sense, since this is the time over which the occupation of state $|\psi^{\rm (st)}_n\rangle$ decays, after which there remains only one excitation in the system. The fast-oscillating components of $\rho_{\alpha\beta\gamma\delta}$ decay even faster, over time $\Gamma^{-1}$.

The signal $R(t)$, Eq.~(\ref{eq:R_probability}), is determined by the two-particle matrix element $\rho_{DD}(t)$. One can show that, in the time range $t\gtrsim\Gamma^{-1}$, the major result of the interaction between the excitation on the two-particle matrix elements is that $\rho_{D1}(t)$ is incremented by $\approx (iJ\Delta/\Gamma)\rho_{D221}(t)$ and the slow part of $\rho_{D2}(t)$ is incremented by $\approx -(J\Delta/\omega_{21}) \rho_{1D21}(t)$. The first of these corrections drops out of the expression for $\rho_{DD}$, whereas the contribution from the second is small compared to $(J/\omega_{21})^2$. Therefore, to the accuracy we are interested in, the interaction between excitations does not affect measurements in the proposed approach.


\begin{acknowledgments}
We are grateful to A. Korotkov and F. Wilhelm for helpful discussions. This work was supported in part by the National Science Foundation
through grants PHY-0555346 and ITR-0085922.
\end{acknowledgments}


\end{document}